\documentstyle[12pt,epsfig]{article}
\begin{document}

\begin{center}
 \Large {\bf  Extracting vector mesons magnetic dipole moment from radiative decays}
\footnote{To appear in the proceedings of the 5th Int.~Conf. ``Quark confinement and the hadron spectrum'', Gargnano, Garda Lake, Italy. 10-14th September 2002}.\\

\vskip.3in
\large{G. Toledo S\'anchez}\\
{\it Instituto de F\'{\i}sica, Universidad Nacional Aut\'onoma de M\'exico \\
A. P. 20-364, Mexico 01000 D.F. M\'exico\\}

\end{center}

\abstract{The possibility that the magnetic dipole moment (MDM) of light charged vector mesons could be measured from radiative processes involving the
production ($\tau \rightarrow \rho \nu \gamma$) and decay ($\rho \rightarrow \pi \pi \gamma$) of vector
mesons is studied in a model independent way, via the soft-photon
approximation. The angular and energy distribution of photons emitted at
small angles respect to the final charged particle is found be sensitive
to the effects of the MDM. We also show that model dependent contributions
have a general structure, by gauge invariance requirements, that allows
suppress them in the same kinematical region where the MDM effect is more
important in the model independent approach.\\
}

The electromagnetic structure of vector mesons (VM) offers an interesting window to strong interaction effects. The electric charge conservation accounts to determine its value for vector mesons. The next non-zero multipole is the magnetic dipole moment (MDM) which is not restricted to a particular value and is matter of study within phenomenological models of QCD \cite{predictions}. However, the experimental determination of none of the vector mesons MDM has been made yet. The reason can be found in the extremely short lifetime of such particles ($\approx 10^{-24}$ s.) and the requirements, for example, of  knowing the initial and final polarizations for the method of spin precession in a magnetic field; rendering to be an impossible task. The same problems are faced by the W gauge boson  whose MDM (here after denoted by $\mu$) is predicted be $\mu=2$ in Bohr's magneton units. Indeed, for VM this value seems to be the favored one to exhibit interesting features  \cite{g2} and  in general is considered as the canonical value.
In the late 60's Zakharov {\it et al.} \cite{zakharov} showed that radiative decays are the only viable option to study resonances, by observing the radiation distribution.
In this work we propose the use of radiative production and/or decay of vector mesons to measure their MDM. First we show, in a model independent way, that the photon angular and energy distribution is sensitive to the MDM, in both radiative decay and production of VM, this is done by choosing a kinematical region to suppress the electric contribution leaving the MDM as the leading one. Then, the electric quadrupole moment (EQM) and  model dependent contributions are shown to be suppressed in this same region.\\
We base our analysis in the fact that the radiative decay amplitude expanded in powers of the photon energy ($\omega$)
$M =\frac{A}{\omega} +B\omega^0 + C\omega +\cdots $ has the two first terms completely determined by the non-radiative process and the electromagnetic properties of the particles involved. This is the so called Low amplitude. The other main ingredient is the electromagnetic vertex $VV\gamma$ \cite{sakita} which contains the information of the electromagnetic structure of the VM. All over the calculations VM are assumed be stable particles. An analysis using a consistently treatment of VM as unstable states requires to compute the full S-matrix amplitude with resonant propagators while keeping gauge invariance \cite{glcPRD00}.
Within our approximations it is possible compute the Low amplitude, and then after sum over polarizations we obtain a typical structure of the form
$ 
\sum_{pols \ i, f} |M|^2 =  \frac{\alpha_2}{\omega^2} sin^2(\theta)+ \alpha_0
\omega^0 + O(\omega)\ ,$
where $\alpha_2$ and $\alpha_0$ are two functions of $O(\omega^0)$ and $\theta$ defines the direction of photon emission with respect to the charged particle in the final state in the decaying particle rest-frame. The MDM dependence is included in $\alpha_0$ and then the $1/\omega^2$ term can be suppressed at small angles to enhance it. This is the basic idea used in both decay and production of VM. For the decay process lets take for example $\rho^+ \rightarrow \pi^+ \pi^0 \gamma$ \cite{glcPRD97}. In Figure 1 we show the energy decay distribution of photons for $\theta=10^0$.
 The short-dashed plot corresponds to the term of $O(\omega^{-2})$ and
the solid, long-dashed long-short-dashed plots are the $O(\omega^0)$ terms when $\mu=1,2$ and $3$ respectively.\\
 Similarly for the production process we can consider the decay
  $\tau \rightarrow V \nu \gamma$ where the effects of MDM \cite{glcPRD99} are enhanced. In both cases the dependence on the value of $\mu$ becomes important even before the middle region.\\
Lets  now consider the higher order contributions, namely the EQM and model dependent terms\cite{glcJPG01}. The total amplitude can be written as: $M \propto \epsilon^{\mu} (L_{\mu} + M_{1,\mu})$. The  first term is
 the electric charge amplitude of $O(\omega^{-1})$, while
$M_{1,\mu}$ is a gauge-invariant one ($k\cdot M_1=0$). The interference  between them and  gauge-invariance imposes it be: $ I \sim -a'_2 sin^2(\theta)$ where $a'_2 $ is a function of any order on $\omega$.
This result holds for the EQM and model dependent contributions,
for example in $\rho \rightarrow \pi\pi\gamma$. In Fig. 2 we plot the angular and energy decay distributions of
photons normalized to the corresponding non-radiative decay rate 
($(1/\Gamma_{nr})d^2\Gamma/dxdy$ with $x \equiv 2\omega/M_\rho$, exhibiting how they get suppressed.\\

\noindent{\bf Acknowledgments}
The author acknowledge partial financial support from Conacyt.

\newpage
\begin{figure}
\vspace{-2in}
\centerline{\epsfig{file=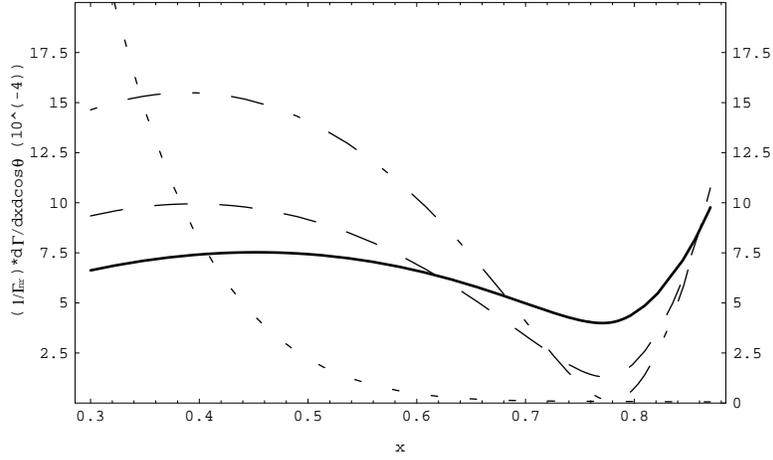,angle=0,width=4in}}
\vskip-1.5in
\caption{Energy decay distribution of photon for $\theta
=10^0$ in the $\rho^+ \rightarrow \pi^+ \pi^0 \gamma$ decay.
 The short-dashed plot corresponds to the term of $O(\omega^{-2})$ and
the solid,
long-dashed long-short-dashed plots are the terms of  $O(\omega^0)$ when
$\mu=1,2$ and $3$ respectively. $x= 2\omega/M_\rho$\label{soft}}
\end{figure}

\begin{figure}
\vspace{-1in}
\centerline{\epsfig{file=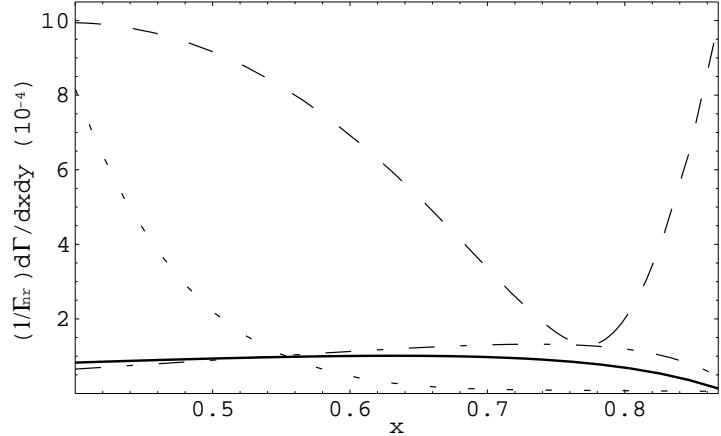,angle=0,width=3.7in}}  
\vspace{.1in}
\caption{Energy decay distributions of photons in the
process $\rho^+ \rightarrow \pi^+\pi^0 \gamma$,  for $\theta=10^0$ ($y
\equiv \cos\theta$). The short-- and long--dashed lines
correspond to the electric charge and the MDM
($\mu =2$)
contributions, respectively. The long-short--dashed and solid lines
correspond to the model-dependent and the EQM
 effects, respectively. \label{model}}
\end{figure}


\begin{thebibliography}{20}
\bibitem{predictions} F.~T.~Hawes and M.~A.~Pichowsky, Phys.~Rev.~C{\bf 59},
 1743(1999); J.~B.~P.~C.~ de Melo and T. Frederico Phys.~Rev.~C{\bf
55}, 2043 (1997).

\bibitem{g2} G. Toledo S\'anchez ,{\it Phys.~Rev.~} {\bf D66}, 097301(2002), and references therein.

\bibitem{zakharov} V.~I.~Zakharov {\it et al.}
Sov.~J.~of Nucl.~Phys. {\bf 8}, 456 (1969).

\bibitem{sakita} B.~Sakita and C.~J.~Goebel, Phys.~Rev.~{\bf 127}, 1787(1962); K. Hagiwara et al. Nucl. Phys. B {\bf 282}, 253(1987).

\bibitem{glcPRD00}G.~L\'opez Castro and G.~Toledo S\'anchez {\it Phys.~Rev.~} 
{\bf D61}, 033007(2000)

\bibitem{glcPRD97} G.~L\'opez Castro and G.~Toledo S\'anchez {\it Phys.~Rev.~} 
{\bf D56}, 4408(1997)

\bibitem{glcPRD99}G.~L\'opez Castro and G.~Toledo S\'anchez {\it Phys.~Rev.~} 
{\bf D60}, 053004(1999)


\bibitem{glcJPG01} G.~L\'opez Castro and G.~Toledo S\'anchez J.~Phys.~G {\bf 27}, 2203(2001)

\end{thebibliography}
\end{document}